\documentclass[aps,prl,twocolumn,superscriptaddress,showpacs]{revtex4-1}
\usepackage{graphicx,hyperref}

\begin{document}


\title{Optical phase cloaking of 700-nm light waves in the far field\\ by a three-dimensional carpet cloak}


\author{Tolga Ergin}
\email[]{tolga.ergin@kit.edu}
\affiliation{Institute of Applied Physics, Karlsruhe Institute of Technology, 76128 Karlsruhe, Germany }
\affiliation{DFG-Center for Functional Nanostructures, Karlsruhe Institute of Technology, 76128 Karlsruhe, Germany }
\author{Joachim Fischer}
\affiliation{Institute of Applied Physics, Karlsruhe Institute of Technology, 76128 Karlsruhe, Germany }
\affiliation{DFG-Center for Functional Nanostructures, Karlsruhe Institute of Technology, 76128 Karlsruhe, Germany }
\author{Martin Wegener}
\homepage[]{http://www.aph.kit.edu/wegener/}
\affiliation{Institute of Applied Physics, Karlsruhe Institute of Technology, 76128 Karlsruhe, Germany }
\affiliation{DFG-Center for Functional Nanostructures, Karlsruhe Institute of Technology, 76128 Karlsruhe, Germany }
\affiliation{Institute of Nanotechnology, Karlsruhe Institute of Technology, 76128 Karlsruhe, Germany }


\date{\today}

\begin{abstract}
Transformation optics is a design tool that connects geometry of space and propagation of light. Invisibility cloaking is a corresponding benchmark example. Recent experiments at optical frequencies have demonstrated cloaking for the light amplitude (``ray cloaking''). In this Letter, we demonstrate far-field cloaking of the light phase (``wave cloaking'') by interferometric microscope-imaging experiments on the previously introduced three-dimensional carpet cloak at 700-nm wavelength and for arbitrary polarization of light.
\end{abstract}

\pacs{42.79.-e,42.79.Ry,81.16.Nd,42.30.Rx}

\maketitle

An object to be made invisible represents a perturbation of a light wave's amplitude and phase with respect to a reference frame. Transformation optics \cite{leonhardt06,pendry06,li08,shalaev08,chen10,leonhardt10} provides a fresh view on optics design in that it allows compensating for these perturbations by using the concepts and mathematics of general relativity. In a broad sense, invisibility cloaking is a particularly demanding example of aberration correction. Several experiments at optical \cite{valentine09,gabrielli09,lee09,renger10,ergin10SC} or even visible \cite{zhang11,chen11,gharghi11,fischer11} frequencies have recently reported successful invisibility cloaking in two \cite{valentine09,gabrielli09,lee09,renger10,zhang11,chen11,gharghi11} and in three \cite{ergin10SC,fischer11,ma10} dimensions, microscopic \cite{valentine09,gabrielli09,lee09,renger10,ergin10SC,fischer11,gharghi11,ma10} or macroscopic \cite{zhang11,chen11} in size, many of which for certain polarizations of light only \cite{valentine09,gabrielli09,lee09,renger10,zhang11,chen11,gharghi11} and fewer in a polarization insensitive manner \cite{ergin10SC,fischer11,ma10}. Importantly, all of these have merely shown invisibility cloaking in the sense of demonstrating reconstructed amplitudes or, equivalently, light intensities.\\ 
Amplitude or ``ray cloaking'' can be achieved in various ways, some of them quite trivial. You can, for example, record the scenery behind the object with a camera and project it onto the object. True cloaking, especially for waves, is a much more subtle undertaking that demonstrates and requires the full potential of transformation optics. In this paper, we experimentally demonstrate optical phase cloaking in the far field, a ``wave cloak'', for what we believe is the first time.\\
\begin{figure}
\includegraphics[width=86mm]{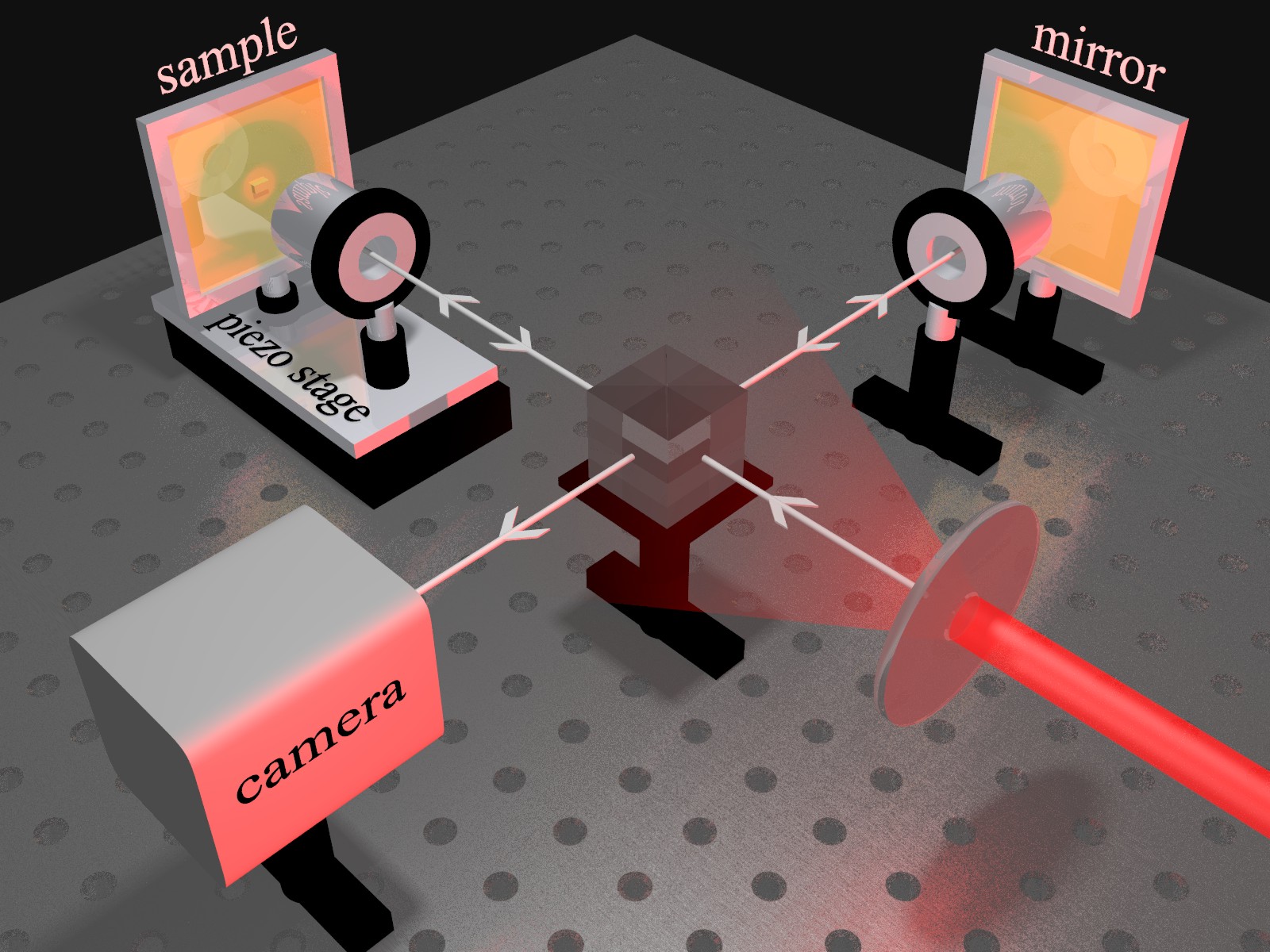}
\caption{\label{setup} Scheme of the experimental setup. A Michelson interferometer is combined with a microscope in each arm. Sample (carpet invisibility cloak) and a flat mirror (serving as a reference frame) form the two end mirrors. Both are illuminated by a laser that is sent through a rotating diffuser, leading to spatially incoherent light.}
\end{figure}
Phase information in an image can be obtained by looking at the object with a microscope (or a telescope) and letting this image interfere with a known reference image. One can, for example, introduce a microscope (or telescope) into each arm of a Michelson interferometer, such that the object's and the reference's image interfere on a screen or video camera. This geometry is illustrated in Fig. \ref{setup}. Scanning the path length of one interferometer arm provides phase sensitivity. To allow for a straightforward interpretation of these data, object and reference must also be illuminated equivalently. This condition can be achieved by illuminating both through the input port of the Michelson interferometer. While coherence is desired in the direction of light propagation (``temporal coherence''), coherence in the lateral direction is quite undesired as it leads to confusing interferences, e.g., from edges. After all, usual microscope or telescope images use spatially incoherent illumination. At the same time, intense illumination is desired to allow for data acquisition in times shorter than typical fluctuations or drifts of the interferometer arms. Here, we meet these illumination requirements by sending a laser (Spectra Physics Inspire) through a rotating diffuser (Thorlabs DG20-120), leading to ``thermal light'' \cite{martienssen64,martienssen66}. We record a movie (roughly 20\,s) of the interferogram while linearly translating the sample arm by a piezo stage. While images recorded with a standing diffuser are completely obscured by speckles, the exposure-time averaging of a single frame with a rotating diffuser leads to rather clean images. If the reference arm is blocked, the interferogram disappears and is replaced by the ordinary amplitude image. In the below experiments, we use microscope objective lenses with a numerical aperture of NA=0.4 (ZEISS LD Achroplan, 20$\times$). Such intermediate NA values allow for the most sensitive test of cloaking as values approaching NA=0 lead to insufficient spatial resolution, whereas for values approaching NA=1 the amplitude contrast diminishes \cite{ergin10OE}. The data shown in this Letter have been taken with linear polarization of light. We find, however, that the orientation of the polarization axis is irrelevant and that identical results are also obtained for circular polarization.\\
\begin{figure}
\includegraphics[width=86mm]{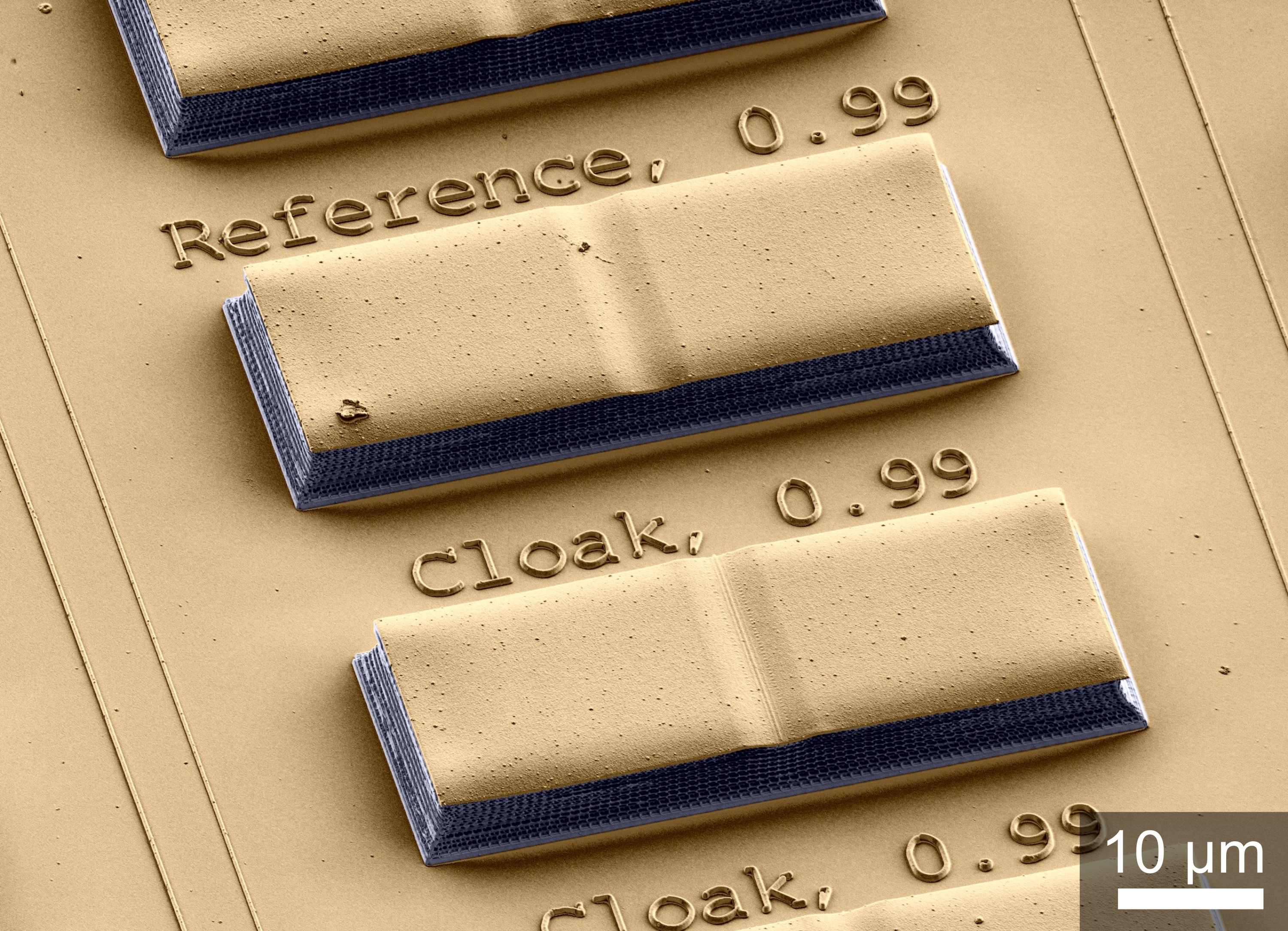}
\caption{\label{REM} Oblique-view electron micrograph of the sample containing a carpet invisibility cloak (bottom) and a reference structure (top). Gold parts are colored in yellow, polymer parts in blue. Taken from Ref. \cite{fischer11}.}
\end{figure}
The visible-frequency carpet-cloak samples used in the present work have been described previously \cite{fischer11} (see Fig. \ref{REM}). We have fabricated them by stimulated-emission-depletion-inspired direct-laser-writing optical lithography \cite{fischer11OME}, followed by coating with a 100-nm thick gold film. In a nutshell, a bump in a metallic mirror (the carpet) allows for hiding objects underneath it. In Fig. \ref{REM}, this structure is shown upside down. The carpet cloak aims at making the bump appear flat, hence unsuspicious. The refractive-index profile required to achieve this task is calculated by transformation optics \cite{li08} and it is mimicked by adjusting the local polymer volume filling fraction, \textit{f}, of a three-dimensional woodpile photonic crystal used in the long-wavelength limit \cite{ergin10SC}. Importantly, we have fabricated a cloak and a reference side by side. The reference is nominally identical to the cloak except that \textit{f}=const., leading to a constant refractive index of about 1.3.\\
\begin{figure}
\includegraphics[width=86mm]{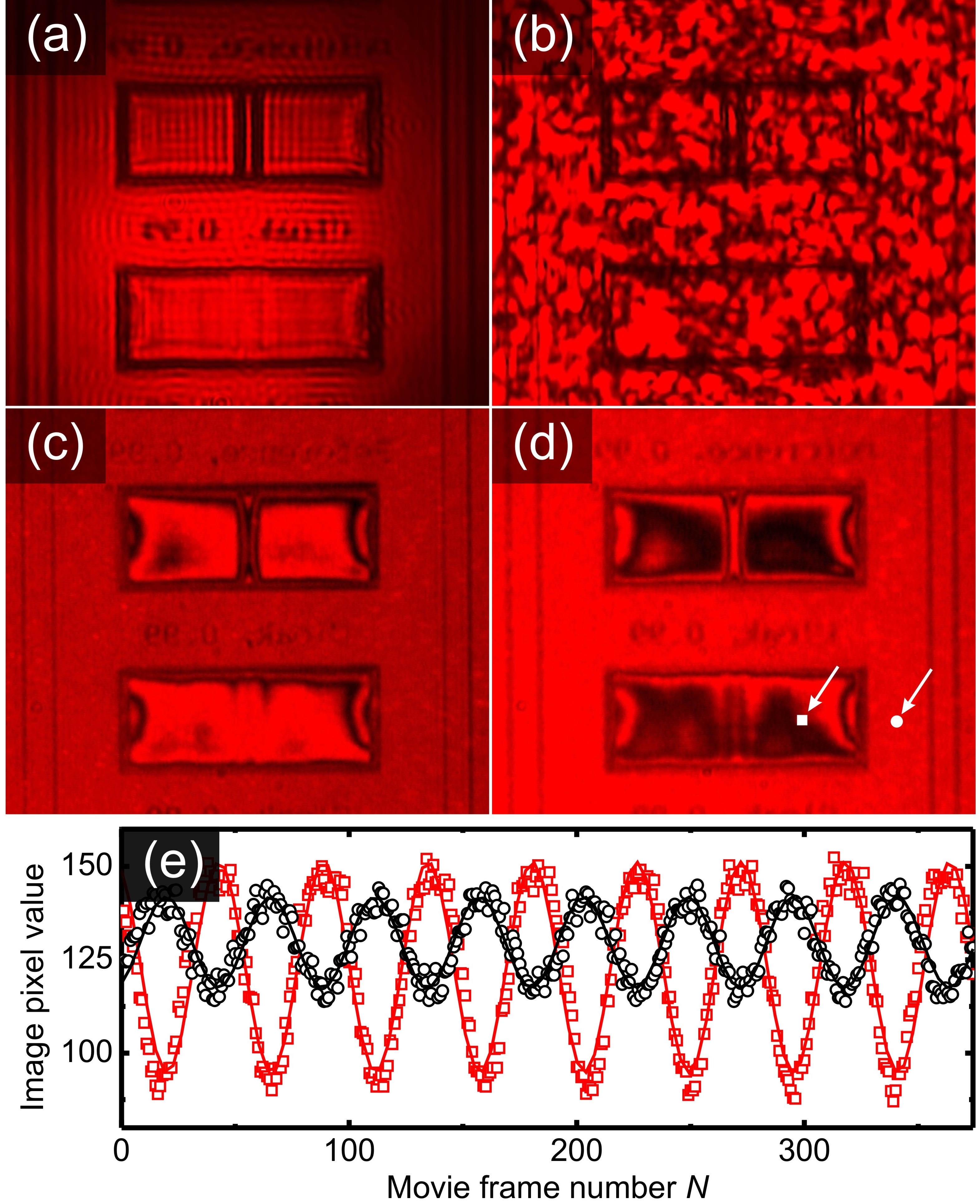}
\caption{\label{speckles} Measured camera images at 700-nm wavelength illumination. (a) Without any diffuser and for blocked reference arm, the images are obscured by interference fringes originating from sample edges. (b) With a standing diffuser in place, the ``spatial coherence'' is effectively reduced, yet the image is now obscured by speckles. In the case of a rapidly rotating diffuser, the camera averages over a large set of such speckle-images during the exposure time of one frame, leading to rather clear and undisturbed images. (c) and (d) show unobscured interference images (i.e., reference arm opened) at two different piezo stage positions, i.e., two different phases. (e) To illustrate the phase reconstruction procedure, raw data (circles and squares) and fits (lines) at two randomly picked pixel positions (marked with white circle and square in (d)) are depicted.}
\end{figure}
Unprocessed raw data taken with the video camera are depicted in Fig. \ref{speckles}. The resulting set of arm-length-dependent images from the camera is still difficult to digest for the observer (see Supplemental Material at [URL will be inserted by publisher] for a sample movie). Intuitive two-dimensional phase images can be extracted in a straightforward manner: It is clear that, for each pixel of the image, the interferometric signal oscillates in a cosine fashion upon linearly scanning one interferometer arm. The local optical phase information is encoded in the phase of that cosine oscillation. Thus, we fit a cosine of the form $a\cdot\cos(\nu\cdot N+\phi)+b$ to each pixel's data (see Fig. \ref{speckles}(e)), where $a$ is the amplitude, $\nu$ is the frequency, $N$ is the movie frame number, $\phi$ is the phase, and $b$ is the offset. Since the frequency has to be constant for all pixels at a fixed wavelength, it is predetermined by fitting the complete cosine-function to several test pixels. The deviations between the extracted frequencies of those test pixels are smaller than $10^{-4}$. For the complete images, the free fit parameters are the amplitude, the phase and the offset. From these fits, the phase is extracted for each pixel, yielding complete two-dimensional phase images. As usual, the phase is only defined modulo 2$\pi$, which leads to many disturbing discontinuous jumps in the resulting raw images. In order to make the images more intuitive to interpret, we shift the branches by a computer algorithm such that continuous two-dimensional phase images result. We find this algorithm to work reliably as long as the phase variation per spatial resolution is substantially smaller than 2$\pi$ and as long as the spatial coherence is sufficiently low.\\ 
\begin{figure}
\includegraphics[width=86mm]{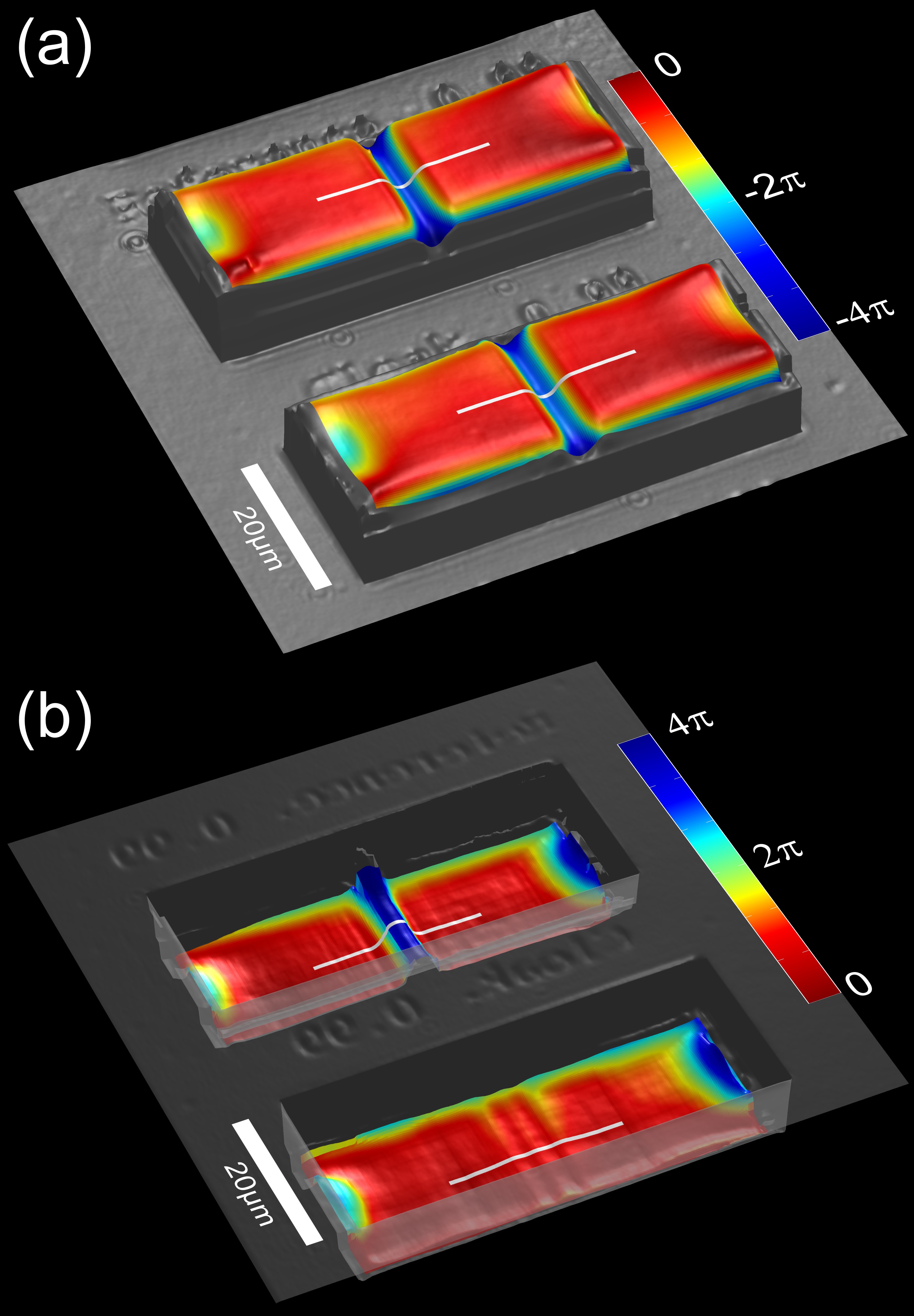}
\caption{\label{phase} Measured phase images. The phase is coded in false color and given in units of rad. The grey colored areas originate from the fact that the phase reconstruction does not work at geometric discontinuities, such as the edges of the structure. Therefore, the phase difference to the colored areas can not be determined. (a) Phase image taken from the air side (compare Fig. \ref{REM}) and (b) taken from the glass-substrate side. The white curves are cuts through the data that are separately shown in Fig. \ref{cuts}.}
\end{figure} 
For consistency, the amplitude images are derived by using the resulting cosine-amplitudes from the fit procedure described above. These amplitude images (not depicted) reproduce our previously published \cite{fischer11} data. As a control experiment, we have measured reference and cloak structure from the air side. The corresponding phase images for an illumination wavelength of 700\,nm are shown in Fig. \ref{phase}(a), where both phase images are equal within the measurement uncertainty for both structures. This observation shows that the two bumps are actually equivalent in terms of width and height (also compare Fig. \ref{REM}). Phase images taken from the glass-substrate side are shown in Fig. \ref{phase}(b). They reveal a striking effect of the cloak. Without cloak, the bump shows up as a phase maximum with a height of more than 2$\pi$. In contrast, with cloak, the phase variation is substantially smaller.\\  
To allow for a more quantitative analysis of these phase images, Fig. \ref{cuts} shows characteristic cuts through these data (see white lines in Figs. \ref{phase}(a) and (b)).  Again, the bumps are identical within the experimental error when inspected from the air side. From the glass-substrate side, the phase maximum is even higher for the reference structure (compared to the air side) because of the average reference refractive index that is larger than unity. For the cloak, the phase maximum is completely gone. Ideally, the phase should be constant. It is presently not clear whether the remaining small wiggles originate from the theoretical design of the carpet cloak or from fabrication imperfections. Regarding amplitude or ray cloaking, the same fabricated structures \cite{fischer11} have recently shown excellent performance in agreement with images obtained by ray tracing calculations (compare Figs. 1(d) and (f) in Ref. \cite{fischer11}). Theoretically calculating the corresponding microscope phase images via solving the Maxwell equations for the cloak and for the entire microscope appears to be out of reach for the near future.\\
\begin{figure}
\includegraphics[width=86mm]{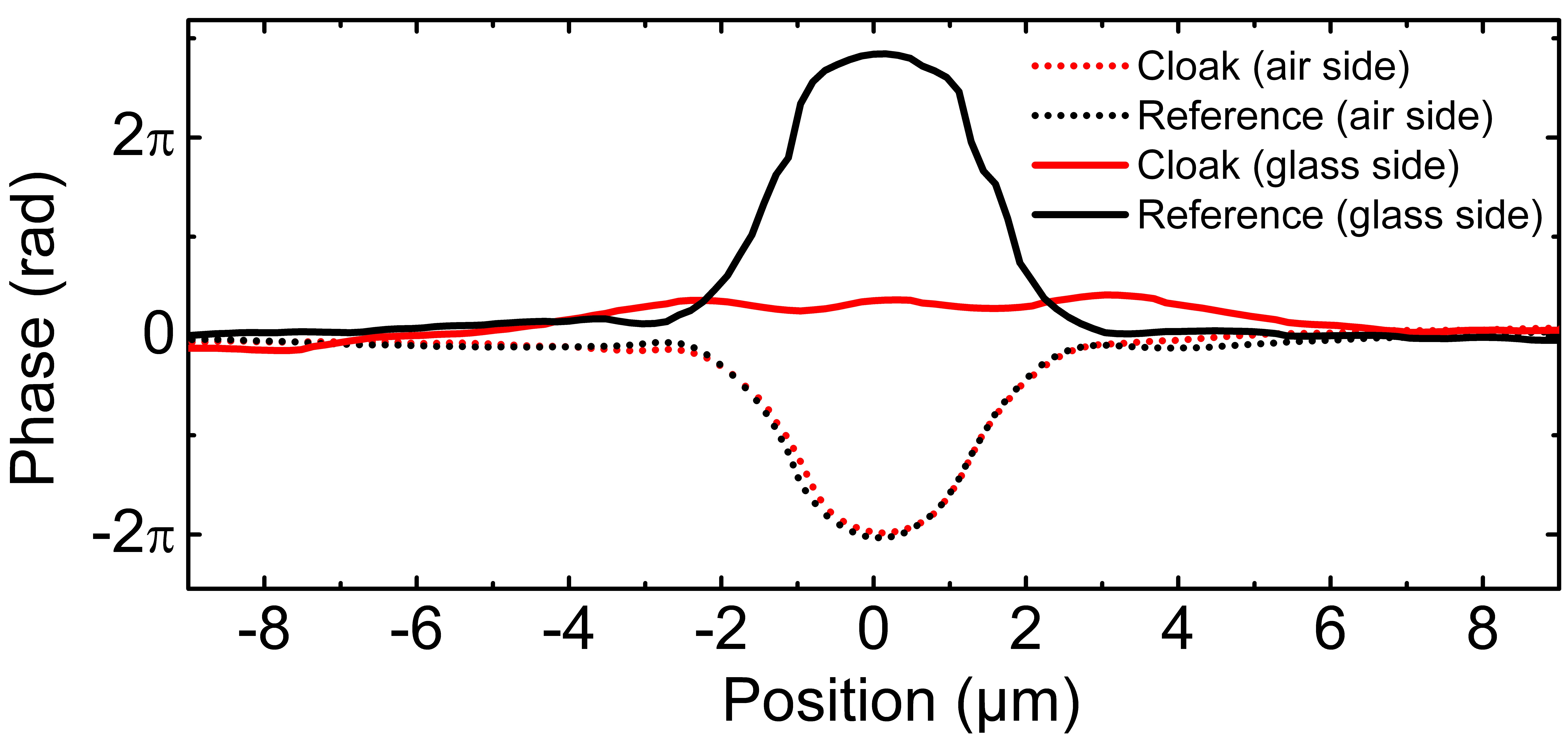}
\caption{\label{cuts} Selected cuts through the phase-image data (see white curves in Fig. \ref{phase}). The solid black (red) curves correspond to measurements on the carpet cloak (reference structure) from the glass-substrate side, the dashed black (red) curves to measurements from the air side.}
\end{figure}
Finally, we have repeated the above interferometric imaging experiment and analysis for illumination laser wavelengths of 750, 800, and 850 nm (not depicted). Clearly, the magnitude of the peak phase changes upon tuning the wavelength. However, if the measured phases are converted into optical path lengths, the measured data at these wavelengths are identical to the ones in Fig. \ref{phase} and \ref{cuts} within the measurement error. Indeed, broadband operation is expected from the design of the dielectric carpet cloak and has previously been demonstrated by us in amplitude-cloaking experiments on the identical structures \cite{fischer11}.\\
In conclusion, we have experimentally shown phase cloaking in the far field at visible frequencies, a ``wave cloak'' rather than just a ``ray cloak'', for what we believe is the first time. Phase sensitivity has been achieved by a Michelson interferometer combined with a microscope and illumination by thermal laser light.  Such wave or phase imaging may also serve as a sensitive test for other devices designed by transformation optics.


%




\begin{acknowledgments}
We thank Nicolas Stenger for discussions. We acknowledge financial support provided by the Deutsche Forschungsgemeinschaft (DFG) and the State of Baden-W\"urttemberg through the DFG-Center for Functional Nanostructures (CFN) within subprojects A1.4 and A1.5. The project PHOME acknowledges the financial support of the Future and Emerging Technologies (FET) programme within the Seventh Framework Programme for Research of the European Commission, under FET-Open grant number 213390. The project METAMAT is supported by the Bundesministerium f\"ur Bildung und Forschung (BMBF). The Ph.D. education of T. Ergin and J. Fischer is embedded in the Karlsruhe School of Optics \& Photonics (KSOP).
\end{acknowledgments}

\bibliography{bibliography}

\providecommand{\noopsort}[1]{}\providecommand{\singleletter}[1]{#1}%
\begin{thebibliography}{20}%
\makeatletter
\providecommand \@ifxundefined [1]{%
 \@ifx{#1\undefined}
}%
\providecommand \@ifnum [1]{%
 \ifnum #1\expandafter \@firstoftwo
 \else \expandafter \@secondoftwo
 \fi
}%
\providecommand \@ifx [1]{%
 \ifx #1\expandafter \@firstoftwo
 \else \expandafter \@secondoftwo
 \fi
}%
\providecommand \natexlab [1]{#1}%
\providecommand \enquote  [1]{``#1''}%
\providecommand \bibnamefont  [1]{#1}%
\providecommand \bibfnamefont [1]{#1}%
\providecommand \citenamefont [1]{#1}%
\providecommand \href@noop [0]{\@secondoftwo}%
\providecommand \href [0]{\begingroup \@sanitize@url \@href}%
\providecommand \@href[1]{\@@startlink{#1}\@@href}%
\providecommand \@@href[1]{\endgroup#1\@@endlink}%
\providecommand \@sanitize@url [0]{\catcode `\\12\catcode `\$12\catcode
  `\&12\catcode `\#12\catcode `\^12\catcode `\_12\catcode `\%12\relax}%
\providecommand \@@startlink[1]{}%
\providecommand \@@endlink[0]{}%
\providecommand \url  [0]{\begingroup\@sanitize@url \@url }%
\providecommand \@url [1]{\endgroup\@href {#1}{\urlprefix }}%
\providecommand \urlprefix  [0]{URL }%
\providecommand \Eprint [0]{\href }%
\providecommand \doibase [0]{http://dx.doi.org/}%
\providecommand \selectlanguage [0]{\@gobble}%
\providecommand \bibinfo  [0]{\@secondoftwo}%
\providecommand \bibfield  [0]{\@secondoftwo}%
\providecommand \translation [1]{[#1]}%
\providecommand \BibitemOpen [0]{}%
\providecommand \bibitemStop [0]{}%
\providecommand \bibitemNoStop [0]{.\EOS\space}%
\providecommand \EOS [0]{\spacefactor3000\relax}%
\providecommand \BibitemShut  [1]{\csname bibitem#1\endcsname}%
\let\auto@bib@innerbib\@empty
\bibitem [{\citenamefont {Leonhardt}(2006)}]{leonhardt06}%
  \BibitemOpen
  \bibfield  {author} {\bibinfo {author} {\bibfnamefont {U.}~\bibnamefont
  {Leonhardt}},\ }\href@noop {} {\bibfield  {journal} {\bibinfo  {journal}
  {Science}\ }\textbf {\bibinfo {volume} {312}},\ \bibinfo {pages} {1777}
  (\bibinfo {year} {2006})}\BibitemShut {NoStop}%
\bibitem [{\citenamefont {Pendry}\ \emph {et~al.}(2006)\citenamefont {Pendry},
  \citenamefont {Schurig},\ and\ \citenamefont {Smith}}]{pendry06}%
  \BibitemOpen
  \bibfield  {author} {\bibinfo {author} {\bibfnamefont {J.~B.}\ \bibnamefont
  {Pendry}}, \bibinfo {author} {\bibfnamefont {D.}~\bibnamefont {Schurig}}, \
  and\ \bibinfo {author} {\bibfnamefont {D.~R.}\ \bibnamefont {Smith}},\
  }\href@noop {} {\bibfield  {journal} {\bibinfo  {journal} {Science}\ }\textbf
  {\bibinfo {volume} {312}},\ \bibinfo {pages} {1780} (\bibinfo {year}
  {2006})}\BibitemShut {NoStop}%
\bibitem [{\citenamefont {Li}\ and\ \citenamefont {Pendry}(2008)}]{li08}%
  \BibitemOpen
  \bibfield  {author} {\bibinfo {author} {\bibfnamefont {J.}~\bibnamefont
  {Li}}\ and\ \bibinfo {author} {\bibfnamefont {J.~B.}\ \bibnamefont
  {Pendry}},\ }\href@noop {} {\bibfield  {journal} {\bibinfo  {journal} {Phys.
  Rev. Lett.}\ }\textbf {\bibinfo {volume} {101}},\ \bibinfo {pages} {203901}
  (\bibinfo {year} {2008})}\BibitemShut {NoStop}%
\bibitem [{\citenamefont {Shalaev}(2008)}]{shalaev08}%
  \BibitemOpen
  \bibfield  {author} {\bibinfo {author} {\bibfnamefont {V.~M.}\ \bibnamefont
  {Shalaev}},\ }\href@noop {} {\bibfield  {journal} {\bibinfo  {journal}
  {Science}\ }\textbf {\bibinfo {volume} {322}},\ \bibinfo {pages} {384}
  (\bibinfo {year} {2008})}\BibitemShut {NoStop}%
\bibitem [{\citenamefont {Chen}\ \emph {et~al.}(2010)\citenamefont {Chen},
  \citenamefont {Chan},\ and\ \citenamefont {Sheng}}]{chen10}%
  \BibitemOpen
  \bibfield  {author} {\bibinfo {author} {\bibfnamefont {H.}~\bibnamefont
  {Chen}}, \bibinfo {author} {\bibfnamefont {C.~T.}\ \bibnamefont {Chan}}, \
  and\ \bibinfo {author} {\bibfnamefont {P.}~\bibnamefont {Sheng}},\
  }\href@noop {} {\bibfield  {journal} {\bibinfo  {journal} {Nature Mater.}\
  }\textbf {\bibinfo {volume} {9}},\ \bibinfo {pages} {387} (\bibinfo {year}
  {2010})}\BibitemShut {NoStop}%
\bibitem [{\citenamefont {Leonhardt}\ and\ \citenamefont
  {Philbin}(2010)}]{leonhardt10}%
  \BibitemOpen
  \bibfield  {author} {\bibinfo {author} {\bibfnamefont {U.}~\bibnamefont
  {Leonhardt}}\ and\ \bibinfo {author} {\bibfnamefont {T.~G.}\ \bibnamefont
  {Philbin}},\ }\href@noop {} {\emph {\bibinfo {title} {Geometry and Light: The
  Science of Invisibility}}}\ (\bibinfo  {publisher} {Dover},\ \bibinfo {year}
  {2010})\BibitemShut {NoStop}%
\bibitem [{\citenamefont {Valentine}\ \emph {et~al.}(2009)\citenamefont
  {Valentine}, \citenamefont {Li}, \citenamefont {Zentgraf}, \citenamefont
  {Bartal},\ and\ \citenamefont {Zhang}}]{valentine09}%
  \BibitemOpen
  \bibfield  {author} {\bibinfo {author} {\bibfnamefont {J.}~\bibnamefont
  {Valentine}}, \bibinfo {author} {\bibfnamefont {J.}~\bibnamefont {Li}},
  \bibinfo {author} {\bibfnamefont {T.}~\bibnamefont {Zentgraf}}, \bibinfo
  {author} {\bibfnamefont {G.}~\bibnamefont {Bartal}}, \ and\ \bibinfo {author}
  {\bibfnamefont {X.}~\bibnamefont {Zhang}},\ }\href@noop {} {\bibfield
  {journal} {\bibinfo  {journal} {Nature Mater.}\ }\textbf {\bibinfo {volume}
  {8}},\ \bibinfo {pages} {568} (\bibinfo {year} {2009})}\BibitemShut {NoStop}%
\bibitem [{\citenamefont {Gabrielli}\ \emph {et~al.}(2009)\citenamefont
  {Gabrielli}, \citenamefont {Cardenas}, \citenamefont {Poitras},\ and\
  \citenamefont {Lipson}}]{gabrielli09}%
  \BibitemOpen
  \bibfield  {author} {\bibinfo {author} {\bibfnamefont {L.~H.}\ \bibnamefont
  {Gabrielli}}, \bibinfo {author} {\bibfnamefont {J.}~\bibnamefont {Cardenas}},
  \bibinfo {author} {\bibfnamefont {C.~B.}\ \bibnamefont {Poitras}}, \ and\
  \bibinfo {author} {\bibfnamefont {M.}~\bibnamefont {Lipson}},\ }\href@noop {}
  {\bibfield  {journal} {\bibinfo  {journal} {Nature Photon.}\ }\textbf
  {\bibinfo {volume} {3}},\ \bibinfo {pages} {461} (\bibinfo {year}
  {2009})}\BibitemShut {NoStop}%
\bibitem [{\citenamefont {Lee}\ \emph {et~al.}(2009)\citenamefont {Lee},
  \citenamefont {Blair}, \citenamefont {Tamma}, \citenamefont {Wu},
  \citenamefont {Rhee}, \citenamefont {Summers},\ and\ \citenamefont
  {Park}}]{lee09}%
  \BibitemOpen
  \bibfield  {author} {\bibinfo {author} {\bibfnamefont {J.~H.}\ \bibnamefont
  {Lee}}, \bibinfo {author} {\bibfnamefont {J.}~\bibnamefont {Blair}}, \bibinfo
  {author} {\bibfnamefont {V.~A.}\ \bibnamefont {Tamma}}, \bibinfo {author}
  {\bibfnamefont {Q.}~\bibnamefont {Wu}}, \bibinfo {author} {\bibfnamefont
  {S.~J.}\ \bibnamefont {Rhee}}, \bibinfo {author} {\bibfnamefont {C.~J.}\
  \bibnamefont {Summers}}, \ and\ \bibinfo {author} {\bibfnamefont
  {W.}~\bibnamefont {Park}},\ }\href@noop {} {\bibfield  {journal} {\bibinfo
  {journal} {Opt. Express}\ }\textbf {\bibinfo {volume} {17}},\ \bibinfo
  {pages} {12922} (\bibinfo {year} {2009})}\BibitemShut {NoStop}%
\bibitem [{\citenamefont {Renger}\ \emph {et~al.}(2010)\citenamefont {Renger},
  \citenamefont {Kadic}, \citenamefont {Dupont}, \citenamefont {Acimovic},
  \citenamefont {Guenneau}, \citenamefont {Quidant},\ and\ \citenamefont
  {Enoch}}]{renger10}%
  \BibitemOpen
  \bibfield  {author} {\bibinfo {author} {\bibfnamefont {J.}~\bibnamefont
  {Renger}}, \bibinfo {author} {\bibfnamefont {M.}~\bibnamefont {Kadic}},
  \bibinfo {author} {\bibfnamefont {G.}~\bibnamefont {Dupont}}, \bibinfo
  {author} {\bibfnamefont {S.~S.}\ \bibnamefont {Acimovic}}, \bibinfo {author}
  {\bibfnamefont {S.}~\bibnamefont {Guenneau}}, \bibinfo {author}
  {\bibfnamefont {R.}~\bibnamefont {Quidant}}, \ and\ \bibinfo {author}
  {\bibfnamefont {S.}~\bibnamefont {Enoch}},\ }\href@noop {} {\bibfield
  {journal} {\bibinfo  {journal} {Opt. Express}\ }\textbf {\bibinfo {volume}
  {18}},\ \bibinfo {pages} {15757} (\bibinfo {year} {2010})}\BibitemShut
  {NoStop}%
\bibitem [{\citenamefont {Ergin}\ \emph
  {et~al.}(2010{\natexlab{a}})\citenamefont {Ergin}, \citenamefont {Stenger},
  \citenamefont {Brenner}, \citenamefont {Pendry},\ and\ \citenamefont
  {Wegener}}]{ergin10SC}%
  \BibitemOpen
  \bibfield  {author} {\bibinfo {author} {\bibfnamefont {T.}~\bibnamefont
  {Ergin}}, \bibinfo {author} {\bibfnamefont {N.}~\bibnamefont {Stenger}},
  \bibinfo {author} {\bibfnamefont {P.}~\bibnamefont {Brenner}}, \bibinfo
  {author} {\bibfnamefont {J.~B.}\ \bibnamefont {Pendry}}, \ and\ \bibinfo
  {author} {\bibfnamefont {M.}~\bibnamefont {Wegener}},\ }\href@noop {}
  {\bibfield  {journal} {\bibinfo  {journal} {Science}\ }\textbf {\bibinfo
  {volume} {328}},\ \bibinfo {pages} {337} (\bibinfo {year}
  {2010}{\natexlab{a}})}\BibitemShut {NoStop}%
\bibitem [{\citenamefont {Zhang}\ \emph {et~al.}(2011)\citenamefont {Zhang},
  \citenamefont {Luo}, \citenamefont {Liu},\ and\ \citenamefont
  {Barbastathis}}]{zhang11}%
  \BibitemOpen
  \bibfield  {author} {\bibinfo {author} {\bibfnamefont {B.}~\bibnamefont
  {Zhang}}, \bibinfo {author} {\bibfnamefont {Y.}~\bibnamefont {Luo}}, \bibinfo
  {author} {\bibfnamefont {X.}~\bibnamefont {Liu}}, \ and\ \bibinfo {author}
  {\bibfnamefont {G.}~\bibnamefont {Barbastathis}},\ }\href@noop {} {\bibfield
  {journal} {\bibinfo  {journal} {Phys. Rev. Lett.}\ }\textbf {\bibinfo
  {volume} {106}},\ \bibinfo {pages} {033901} (\bibinfo {year}
  {2011})}\BibitemShut {NoStop}%
\bibitem [{\citenamefont {Chen}\ \emph {et~al.}(2011)\citenamefont {Chen},
  \citenamefont {Luo}, \citenamefont {Zhang}, \citenamefont {Jiang},
  \citenamefont {Pendry},\ and\ \citenamefont {Zhang}}]{chen11}%
  \BibitemOpen
  \bibfield  {author} {\bibinfo {author} {\bibfnamefont {X.}~\bibnamefont
  {Chen}}, \bibinfo {author} {\bibfnamefont {Y.}~\bibnamefont {Luo}}, \bibinfo
  {author} {\bibfnamefont {J.}~\bibnamefont {Zhang}}, \bibinfo {author}
  {\bibfnamefont {K.}~\bibnamefont {Jiang}}, \bibinfo {author} {\bibfnamefont
  {J.~B.}\ \bibnamefont {Pendry}}, \ and\ \bibinfo {author} {\bibfnamefont
  {S.}~\bibnamefont {Zhang}},\ }\href@noop {} {\bibfield  {journal} {\bibinfo
  {journal} {Nat. Commun.}\ }\textbf {\bibinfo {volume} {2}},\ \bibinfo {pages}
  {176} (\bibinfo {year} {2011})}\BibitemShut {NoStop}%
\bibitem [{\citenamefont {Gharghi}\ \emph {et~al.}(2011)\citenamefont
  {Gharghi}, \citenamefont {Gladden}, \citenamefont {Zentgraf}, \citenamefont
  {Liu}, \citenamefont {Yin}, \citenamefont {Valentine},\ and\ \citenamefont
  {Zhang}}]{gharghi11}%
  \BibitemOpen
  \bibfield  {author} {\bibinfo {author} {\bibfnamefont {M.}~\bibnamefont
  {Gharghi}}, \bibinfo {author} {\bibfnamefont {C.}~\bibnamefont {Gladden}},
  \bibinfo {author} {\bibfnamefont {T.}~\bibnamefont {Zentgraf}}, \bibinfo
  {author} {\bibfnamefont {Y.}~\bibnamefont {Liu}}, \bibinfo {author}
  {\bibfnamefont {X.}~\bibnamefont {Yin}}, \bibinfo {author} {\bibfnamefont
  {J.}~\bibnamefont {Valentine}}, \ and\ \bibinfo {author} {\bibfnamefont
  {X.}~\bibnamefont {Zhang}},\ }\href@noop {} {\bibfield  {journal} {\bibinfo
  {journal} {Nano Lett.}\ } (\bibinfo {year} {2011})},\ \bibinfo {note}
  {(published online)}\BibitemShut {NoStop}%
\bibitem [{\citenamefont {Fischer}\ \emph {et~al.}(2011)\citenamefont
  {Fischer}, \citenamefont {Ergin},\ and\ \citenamefont {Wegener}}]{fischer11}%
  \BibitemOpen
  \bibfield  {author} {\bibinfo {author} {\bibfnamefont {J.}~\bibnamefont
  {Fischer}}, \bibinfo {author} {\bibfnamefont {T.}~\bibnamefont {Ergin}}, \
  and\ \bibinfo {author} {\bibfnamefont {M.}~\bibnamefont {Wegener}},\
  }\href@noop {} {\bibfield  {journal} {\bibinfo  {journal} {Opt. Lett.}\
  }\textbf {\bibinfo {volume} {36}},\ \bibinfo {pages} {2059} (\bibinfo {year}
  {2011})}\BibitemShut {NoStop}%
\bibitem [{\citenamefont {Ma}\ and\ \citenamefont {Cui}(2010)}]{ma10}%
  \BibitemOpen
  \bibfield  {author} {\bibinfo {author} {\bibfnamefont {H.~F.}\ \bibnamefont
  {Ma}}\ and\ \bibinfo {author} {\bibfnamefont {T.~J.}\ \bibnamefont {Cui}},\
  }\href@noop {} {\bibfield  {journal} {\bibinfo  {journal} {Nat. Commun.}\
  }\textbf {\bibinfo {volume} {1}},\ \bibinfo {pages} {21} (\bibinfo {year}
  {2010})}\BibitemShut {NoStop}%
\bibitem [{\citenamefont {Martienssen}\ and\ \citenamefont
  {Spiller}(1964)}]{martienssen64}%
  \BibitemOpen
  \bibfield  {author} {\bibinfo {author} {\bibfnamefont {W.}~\bibnamefont
  {Martienssen}}\ and\ \bibinfo {author} {\bibfnamefont {E.}~\bibnamefont
  {Spiller}},\ }\href@noop {} {\bibfield  {journal} {\bibinfo  {journal} {Am.
  J. Phys.}\ }\textbf {\bibinfo {volume} {32}},\ \bibinfo {pages} {919}
  (\bibinfo {year} {1964})}\BibitemShut {NoStop}%
\bibitem [{\citenamefont {Martienssen}\ and\ \citenamefont
  {Spiller}(1966)}]{martienssen66}%
  \BibitemOpen
  \bibfield  {author} {\bibinfo {author} {\bibfnamefont {W.}~\bibnamefont
  {Martienssen}}\ and\ \bibinfo {author} {\bibfnamefont {E.}~\bibnamefont
  {Spiller}},\ }\href@noop {} {\bibfield  {journal} {\bibinfo  {journal} {Phys.
  Rev. Lett.}\ }\textbf {\bibinfo {volume} {16}},\ \bibinfo {pages} {531}
  (\bibinfo {year} {1966})}\BibitemShut {NoStop}%
\bibitem [{\citenamefont {Ergin}\ \emph
  {et~al.}(2010{\natexlab{b}})\citenamefont {Ergin}, \citenamefont {Halimeh},
  \citenamefont {Stenger},\ and\ \citenamefont {Wegener}}]{ergin10OE}%
  \BibitemOpen
  \bibfield  {author} {\bibinfo {author} {\bibfnamefont {T.}~\bibnamefont
  {Ergin}}, \bibinfo {author} {\bibfnamefont {J.~C.}\ \bibnamefont {Halimeh}},
  \bibinfo {author} {\bibfnamefont {N.}~\bibnamefont {Stenger}}, \ and\
  \bibinfo {author} {\bibfnamefont {M.}~\bibnamefont {Wegener}},\ }\href@noop
  {} {\bibfield  {journal} {\bibinfo  {journal} {Opt. Express}\ }\textbf
  {\bibinfo {volume} {18}},\ \bibinfo {pages} {20535} (\bibinfo {year}
  {2010}{\natexlab{b}})}\BibitemShut {NoStop}%
\bibitem [{\citenamefont {Fischer}\ and\ \citenamefont
  {Wegener}(2011)}]{fischer11OME}%
  \BibitemOpen
  \bibfield  {author} {\bibinfo {author} {\bibfnamefont {J.}~\bibnamefont
  {Fischer}}\ and\ \bibinfo {author} {\bibfnamefont {M.}~\bibnamefont
  {Wegener}},\ }\href@noop {} {} (\bibinfo {year} {2011}),\ \bibinfo {note}
  {{Opt. Mater. Express} (accepted)}\BibitemShut {NoStop}%
\end{thebibliography}%

\end{document}